# Entropy driven stability of chiral single-walled carbon nanotubes[*]

Yann Magnin[1,3], Hakim Amara[2], François Ducastelle[2], Annick Loiseau[2], Christophe Bichara[1,*]

[1] Aix Marseille Univ, CNRS, Centre Interdisciplinaire de Nanoscience de Marseille, Campus de Luminy, Case 913, F-13288, Marseille, France

[2] Laboratoire d'Etude des Microstructures, ONERA-CNRS, UMR104, Université Paris-Saclay, BP 72, 92322 Châtillon Cedex, France

[3] Present address: MultiScale Material Science for Energy and Environment, MIT-CNRS Joint Laboratory at MIT, Cambridge, Massachusetts 02139, USA.

*Correspondence to: bichara@cinam.univ-mrs.fr.

**Abstract:**

Single-walled carbon nanotubes are hollow cylinders, that can grow centimeters long by carbon incorporation at the interface with a catalyst. They display semi-conducting or metallic characteristics, depending on their helicity, that is determined during their growth. To support the quest for a selective synthesis, we develop a thermodynamic model, that relates the tube-catalyst interfacial energies, temperature, and the resulting tube chirality. We show that nanotubes can grow chiral because of the configurational entropy of their nanometer-sized edge, thus explaining experimentally observed temperature evolutions of chiral distributions. Taking the chemical nature of the catalyst into account through interfacial energies, structural maps and phase diagrams are derived, that will guide a rational choice of a catalyst and growth parameters towards a better selectivity.

**One Sentence Summary:**
Modeling the interface of carbon nanotubes with their seeding catalyst nanoparticle reveals the origin of their chirality.

Electronic properties of single walled carbon nanotubes (SWNTs) depend on the way they are rolled along their axis - their "chirality" -, characterized by two indices $(n, m)$. Controlling chirality during the tube's synthesis would avoid costly sorting, and trigger the implementation of promising applications, such as the use of SWNT yarns as strong, light and conductive wires *(1)*, or the development of SWNT-based electronics *(2)*, with the ultimate goal of overcoming the limitations of silicon. Significant breakthroughs have been reported *(3, 4)* and progress towards "Carbon Nanotube computers" *(5,6)* has been very rapid. However, selective synthesis still appears to be the weak link, though new experiments using solid state catalysts *(7,8,9)* have reported a chiral specific growth of SWNTs. Detailed mechanisms underlying this selective growth are still debated, thus underlining the need for realistic growth models explicitly including the role of the catalyst. Existing models either focus on kinetics *(10)* , neglecting the role of the catalyst *(11, 12)*, but fail to calculate chiral distributions in line with experiments. Atomistic computer simulations emphasize on chemical accuracy *(13, 14)*, but need to be complemented with a model so as to provide a global understanding of the process. Here, we develop a thermodynamic modelling of the interface between the tube and the catalyst, to relate its properties to the resulting chiral distribution obtained during chemical vapor deposition (CVD) synthesis experiments.

Vapor-liquid-solid and Vapor-solid-solid CVD processes have both been used to grow SWNTs *(8)*, the latter leading to a $(n, m)$ selectivity. Growth can proceed through tangential or perpendicular modes *(15)*, and ways to control them have been proposed recently *(16)*. For specific catalysts and growth conditions favoring the perpendicular mode, a pronounced near armchair selectivity can be observed *(16)*. In such a mode, the interface between the tube and the catalyst nanoparticle (NP) is limited to a line, and a simple model describing the thermodynamic stability of the tube-nanoparticle system can be developed. We thus consider an ensemble of configurations of a catalyst NP, possibly a metal or a carbide, in perpendicular contact with a $(n, m)$ SWNT, as in Fig.1. The total numbers of carbon and catalyst atoms are constant. Configurations differ by the structure of the NP-tube interface, defined by $(n, m)$, for which we have $(n + m)$ SWNT-NP bonds, with typically $10 < n + m < 50$. On the tube edge, $2\,m$ among them are armchair, and $(n - m)$ zigzag *(17)*. In a first approximation, the atomic structure of the NP is neglected, and the catalyst appears as a smooth flat surface, in a jellium-like approximation. The interface is then a simple closed loop with two kinds of species: armchair and zigzag contact atoms. Under these conditions, the total energy of the system can be split in three terms:

$$E(n, m) = E_0 + E_{curv}(n, m) + E_{Int}(n,m) \qquad (1)$$

Where $E_0$ includes all terms independent of $(n, m)$, such as the energy of the 3-fold coordinated carbon atoms in the tube wall, and the atoms forming the NP. The surface energy of the NP and the very weak surface energy of the tube are also included in $E_0$, because these surfaces are kept constant. Note that this model could possibly apply also in tangential mode, if the lateral tube/catalyst interaction does not depend on $(n, m)$.

The $(n, m)$ dependent energy terms concern the tube curvature, and its interface with the NP. Using Density Functional Theory (DFT) calculations, Gülseren *et al.* *(18)* evaluated the curvature energy of the isolated tube as $E_{Curv} = 4\,\alpha\,D_{CNT}^{-2}$, $D_{CNT}$ being the tube diameter, and $\alpha = 2.14\,eV\text{Å}^2$ per C atom. We assume that the interfacial energy for a $(n, m)$ tube in contact

with the NP surface depends only on the number of its $2m$ armchair and $(n-m)$ zigzag contacts:

$$E_{Int}^{(n,m)} = 2m\, E_{Int}^A + (n-m)\, E_{Int}^Z \qquad (2)$$

where the armchair ($E_{Int}^A$) and zigzag ($E_{Int}^Z$) interfacial energies are given by $E_{Int}^X = \gamma_G^X + E_{Adh}^X$, with $X$ standing for $A$ or $Z$. The edge energy per dangling bond, $\gamma_G^X$, is positive since it is the energy cost of cutting a tube or a graphene ribbon, and depends on the type of edge created. The adhesion energy of the tube in contact with the NP, $E_{Adh}^X$, is negative since energy is gained by reconnecting a cut tube to the NP. $E_{Int}^X$, sum of these two terms has to be positive to create a driving force for SWNT formation. DFT calculations of the edge energies of different edge configurations of a (8,4) tube (Fig. 2), show no preferential ordering. We thus assume that all tube catalyst interfaces with the same number of armchair and zigzag contacts have the same energy.

This leads us to introduce the edge configurational entropy as a central piece of the model. We assume that the tube is cut almost perpendicular to its axis, forming the shortest possible interface, for a given $(n,m)$. We neglect vibrational entropy contributions, that are essentially the same for all tubes, except for radial breathing modes. Armchair 2-fold coordinated C atoms always come as a pair, thus this entropy that relates the number of ways of putting $(n-m)$ zigzag C atoms and $m$ pairs of armchair atoms on $n$ sites (degeneracy) is:

$$\frac{S(n,m)}{k_B} = \ln \frac{n!}{m!(n-m)!} \qquad (3)$$

Interfacial energies can be evaluated using DFT calculations, described in Methods. In agreement with *(19, 17)*, we find $\gamma_G^A = 2.06$ eV / bond and $\gamma_G^Z = 3.17$ eV / bond for graphene, and 1.99 and 3.12 respectively for cutting (6,6) and (12,0) tubes. The lower value of $\gamma_G^A$ is due to the relaxation (shortening) of the C-C bonds of the armchair edge that stabilizes it. Adhesion energies of (10,0) and (5,5) tubes on icosahedral clusters of various metals, including Fe, Co, Ni, Cu, Pd and Au were calculated in *(20, 21)*. Thus, orders of magnitude for interface energies, $E_{Int}^X$, of armchair and zigzag terminations in contact with typical catalysts can be estimated: they lie between 0.0 and 0.5 eV / bond, with $E_{Int}^A < E_{Int}^Z$ for these metals. An example of free energy, and corresponding probability distribution is plotted as a function of $(n,m)$ in Figure S1.

Instead of focusing on a specific catalytic system, it is more relevant at this stage to study the general properties of the model that links the $(n,m)$ indexes of a SWNT, to three parameters characterizing its CVD growth conditions, namely temperature and the interfacial energies of armchair ($E_{Int}^A$) and zigzag ($E_{Int}^Z$) tube-catalyst contacts. For each set of parameters, a free energy can be calculated, and its minimization yields the stable $(n,m)$ value. This model displays similarities with a simple alloy model on a linear chair, but the curvature term, dominant for small diameters, and the small and discrete values of $n$ and $m$, prevent to make it analytically solvable, except for ground states, i.e. stable structures at zero Kelvin, for which a solution is provided in Methods. We thus define a 3-dimensional space of stable configurations in the $(T, E_{Int}^A, E_{Int}^Z)$ coordinates.

Setting $T$, and hence the entropy contribution to zero, the ground states are readily calculated and displayed in Fig. 3-A. Only armchair or zigzag tubes are found to be stable, separated by a line $E_{Int}^Z = \frac{4}{3} E_{Int}^A$. With increasing temperature, they become unstable, and a transition towards chiral tubes takes place. Fig. 3-B is a contour plot of the surface defined by the transition temperatures. Above this surface, for each set of $(T, E_{Int}^A, E_{Int}^Z)$ parameters, a chiral $(n, m)$ tube is found stable, defining "volumes" of stability for each chirality. To explore it, we can cut slices at constant temperature to obtain an isothermal stability map (Fig. 3-C). In such maps, only the most stable $(n, m)$ tube structures are shown, while the model yields a distribution of chiralities for each $(T, E_{Int}^A, E_{Int}^Z)$ point. Within a $(n, m)$ domain, this distribution is not constant, especially close to the boundaries, that are calculated by searching for points where the free energies, and hence the probabilities of two competing structures are equal. As illustrated in Fig. S1-B, around the chirality that displays a maximal probability set to 1, neighboring chiralities have non negligible contributions, that depend on $(T, E_{Int}^A, E_{Int}^Z)$. We can also fix either $E_{Int}^A$ or $E_{Int}^Z$ to obtain temperature dependent "phase diagrams", as in Fig. 4-A (for $E_{Int}^A = 0.15$ eV / bond) and Fig. 4-B (for $E_{Int}^Z = 0.25$ eV / bond). As an example, we can follow the temperature stability of a (6,6) tube. Fig. 4-A shows a large stability range with a maximal stability temperature rising from 200 to 800 K by increasing $E_{Int}^Z$ from 0.20 to 0.30 eV / bond, whereas the second map, orthogonal to the first one in the 3-d configuration space, shows an upper temperature limit varying from 500 to 700 K, within a narrower $E_{Int}^A$ range. Above the armchair tubes, chiral *(n, n-m)* tubes become stable starting with *(n, n-1)*, and then with increasing $(n - m)$ values such as (6,5), (7,5), etc. Chiral tubes, i.e. tubes different from armchair or zigzag, are stabilized at finite temperature by the configurational entropy of the tube edge.

An isothermal map calculated at 1000 K is plotted in Fig. 3-C. Chiral tubes are spread along the $E_{Int}^Z = \frac{4}{3} E_{Int}^A$ diagonal, between armchair and zigzag ones. Small diameter tubes are stabilized for larger values of $(E_{Int}^A, E_{Int}^Z)$, hence for weaker adhesion energies of the tube on the catalyst. Larger diameter tubes are obtained for small values of $(E_{Int}^A, E_{Int}^Z)$, because the entropy cannot counterbalance the energy cost of the interface, proportional to $n + m$. A comparison of maps at 1000 and 1400 K is given in Figure S3. As shown in Movie S1, the effect of increasing temperature is to expand and shift the stability domain of chiral tubes along and on both sides of the $E_{Int}^Z = \frac{4}{3} E_{Int}^A$ diagonal, with a larger spread on the armchair side. The stability domain of chiralities between central $(2n, n)$ and close to armchair $(n, n - 1)$ expands significantly at high temperature. We note however that the free energy differences become smaller, leading to broader chiral distributions, thus explaining the lack of selectivity reported for tubes grown at very high temperature by electric arc or laser ablation methods *(22)*.

This very simple model displays a fair agreement with literature data, as illustrated in the following examples. Fig. 3-B suggests a way to grow either zigzag or armchair tubes, the latter being metallic for any diameter. For both, growth kinetics is slow, because each new ring of carbon atoms has to nucleate, once the previous one has been completed *(11, 12)*. To overcome this nucleation barrier, one should seek regions in the map, where such tubes remain stable at high temperature. For armchair species, this corresponds to the lower right corner of the map in Fig. 3-B, where the adhesion energy of armchair edges is strong, and that of zigzag ones is weak, and the opposite for the interfacial energies. Such requirements have possibly been met in high temperature (1473 K) CVD experiments *(23)*, that also used thiophene in the feedstock. Those experiments might indicate that the presence of sulfur at the interface could modify the relative interaction strength of zigzag and armchair edges with the Fe NP.

The temperature dependence of the chiral distributions, measured by photoluminescence in *(24, 25, 26)*, Raman and transmission electron spectroscopies *(27)* seems more robust. The maps presented in Fig. 4-A and Fig. 4-B are consistent with these experiments showing that armchair or near armchair chiralities - (6,6) and (6,5) - are grown at low temperature (873 K), and that the chiral distribution gradually shifts towards larger chiral angles - (7,5), (7,6) and (8,4) ...- at higher temperatures. Referring to our model, this suggests that Co and Fe based catalysts used in these experiments correspond to interfacial energy values around $E_{Int}^A = 0.15$ eV / bond and $E_{Int}^Z = 0.24$ eV / bond, as indicated by the dashed boxes in the maps. A quantitative comparison with four different sets of experimental data is given in Fig. S2, showing a slight tendency to overestimate the width of the distributions. This overestimation partly results from the fact that we use a two-parameter thermodynamic model to account for experiments that include the variability in the catalyst size and chemical composition and the growth kinetics. Our results also confirm that overlooking metallic tubes in photoluminescence experiments introduces a serious bias in the resulting chiral distribution. Further, the dependence of the quantum yield to chiral angles of semi-conducting tubes may also contribute to underestimate the width of the experimental distributions.

The present model thus sets a framework for understanding why a number of experiments, using metallic catalysts in perpendicular growth conditions, as discussed in *(16)*, report a near armchair selectivity. For such catalysts, $E_{Int}^A$ is generally lower than $E_{Int}^Z$ *(20)*. At low temperature, zigzag or armchair tubes are thermodynamically favored, but may not always be obtained due to kinetic reasons. On the armchair side, our model indicates that close to armchair helicities are then favored by a temperature increase, because their stability domain is large, and they are less kinetically impaired *(11)*. At even higher temperatures, tube chiralities tending towards *(2n, n)* indexes should be stabilized by their larger edge configurational entropy, but their stability domains turn out to be narrower in the present model. Taking the atomic structure of the catalyst into account in our model could rule out some neighboring structures, and contribute to open up these domains.

Concerning the practical use of these maps, a first issue is to select the appropriate location for a catalyst in the $(T, E_{Int}^A, E_{Int}^Z)$ coordinates, so as to favor the desired tube helicity. Looking at Fig. 3-C, one can see that the largest and most interesting parameter ranges correspond to either metallic armchair tubes, or to $(n, n-1)$ and $(n, n-2)$ semi-conducting tubes. A second, more difficult issue is to design a catalyst that would display appropriate $E_{Int}^A$ and $E_{Int}^Z$ values. DFT-based calculations, in the same spirit as those in *(20, 21, 7, 28, 9)* should probably be helpful. However, the evidence of the important role of the edge configurational entropy questions the possibility to explain the high selectivity reported in *(7, 9)* on the basis of a structural or symmetry matching. The intrinsic disorder at the edge could be taken into account by averaging over various atomic configurations, and using Molecular Dynamics at finite temperature.

The present model reevaluates the role of thermodynamics in the understanding of SWNT growth mechanisms. It accounts for experimental evidence, such as the close to armchair preferential selectivity, hitherto attributed to kinetics *(11)*, and the temperature dependent trends in chiralities. It also provides a guide to design better, more selective catalysts. One should however bear in mind the importance of kinetics in a global understanding of the SWNT growth process. An attempt to combine thermodynamic and kinetic aspects of the growth has been proposed in *(12)*, but, overlooking the role of the edge configurational entropy, it led to unrealistic chiral distributions. Those resulting from the present thermodynamic analysis are slightly broader than the experimental ones (Fig. S2), but should be narrower, if the reported chirality dependence of the growth kinetics *(11)* were taken into account. Due to the high synthesis temperatures, and the very small size of the interface, there might be SWNT growth regimes where the atomic mobility and the residence time of atoms close to the interface are large enough to achieve a local thermodynamic equilibrium.

**Acknowledgments:**

**Funding:** Support from the French research funding agency (ANR), under grant 13-BS10-0015-01 (SYNAPSE) is gratefully acknowledged.

**Author contributions:**
CB designed the project, YM, HA, FD and CB developed the model, all authors contributed to the data analysis and the manuscript preparation. CB thanks Prof. Pierre Müller for stimulating discussions.

**Competing interests:** Authors declare no competing interests.

**Data and materials availability:** All data is available in the main text or the Supplementary Materials.

**Supplementary Materials:**

Methods
Figs. S1 to S3

**Other Supplementary Materials for this manuscript include the following:**

Movies S1

Captions for Movie S1

Data S1: Computer codes: Chirality_Map.f90, Chirality_Map.py

Captions for Data S1

Data S2: Numerical data of Fig. S2

Captions for Data S2

# Figures

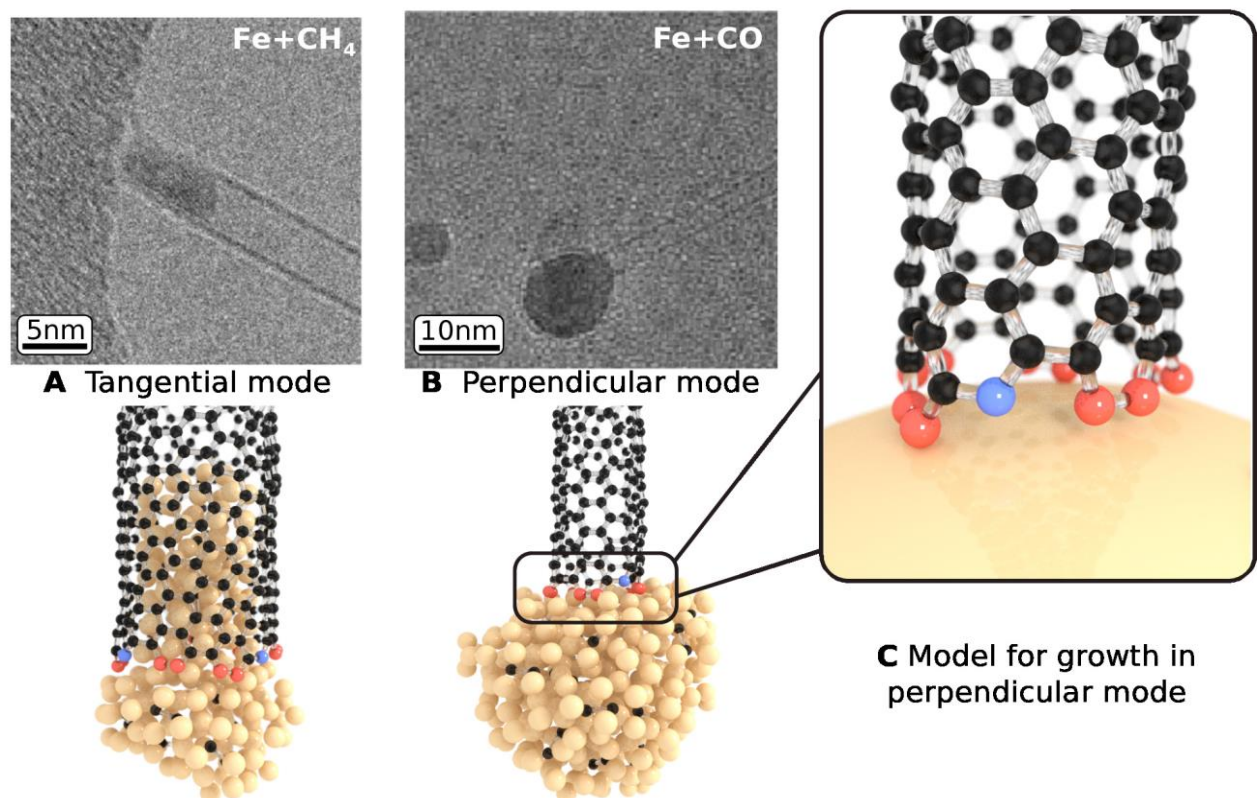

**Fig. 1. From experiments to a model**. **Top:** post-synthesis TEM images of a SWNT attached to the nanoparticle from which it grew at 1073 K, using either $CH_4$ (**A**), or CO (**B**) carbon feedstocks, leading to tangential or perpendicular growth mode illustrated at the atomic scale in the bottom line. The experiments, and our analysis of the growth modes are described in *(16)*. Sketch of the model (**C**), with a SWNT in perpendicular contact with a structureless catalyst. Armchair edge atoms are in red, zigzag ones in blue.

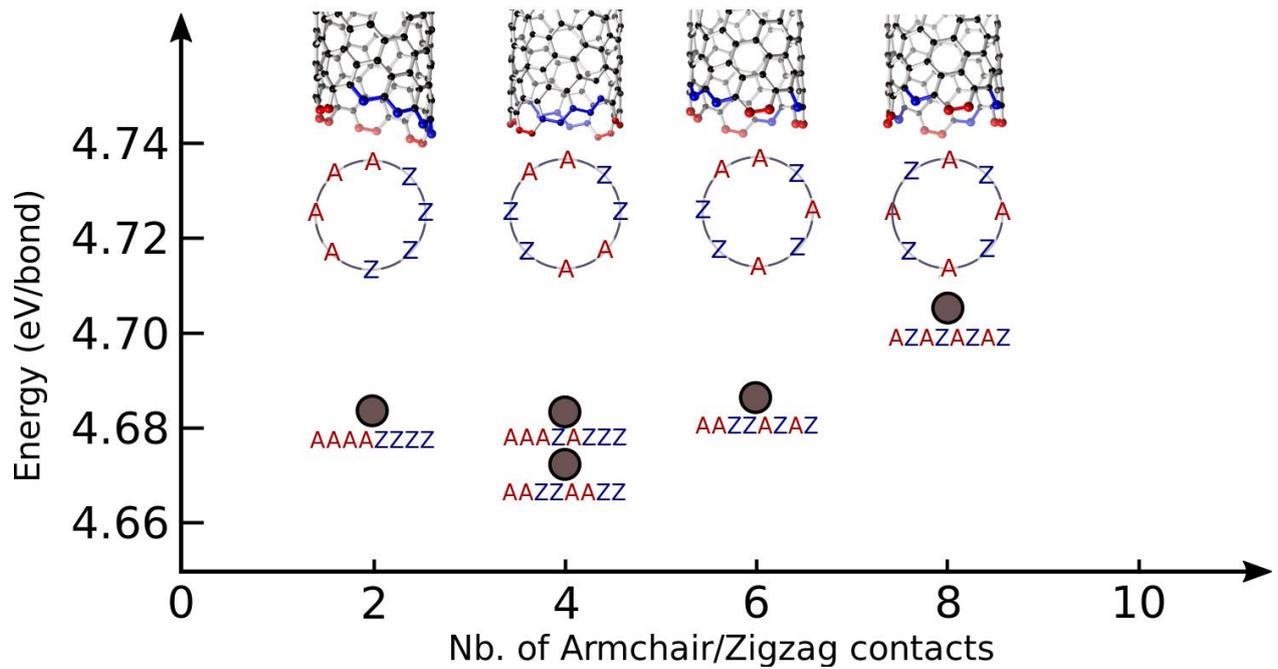

**Fig. 2. Key elements of the model**. **Top**: Different ways of cutting a (8,4) tube, leading to the formation of zigzag (blue) and armchair (red) under-coordinated atoms. For a (8,4) tube, there are 70 different edge configurations with almost the same energy. **Bottom**: Formation energies of all possible (8, 4) edges, from DFT calculations described in Methods. They lie within 25 meV / bond, and can thus be considered degenerate.

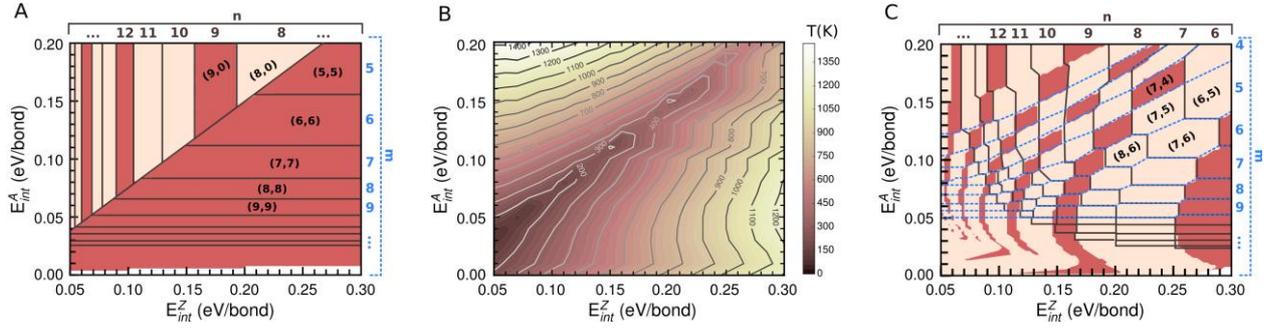

**Fig. 3. Structural maps**. **A)** Map of the ground states, with armchair tubes in the lower right corner, and zigzag ones in the upper left corner, separated by a line $E_{Int}^Z = \frac{4}{3} E_{Int}^A$. Small diameter tubes, e.g. (5,5) and (8,0), are obtained for large values of the interfacial energies $E_{Int}$, while stability domains of large diameter tubes are narrower, with a width decaying as $\frac{1}{n(n+1)}$, and obtained for small values of $E_{Int}$. **B)** Contour plot of the highest temperatures of stability of the ground state structures, armchair or zigzag. Chiral tubes are only found above this surface, stabilized by the configurational entropy of the tube's edge. Note that armchair and zigzag tubes can remain stable at high temperatures, in the bottom right and upper left corners respectively. **C)** Chirality map at 1000 K. Iso-$n$ (resp. iso-$m$) values are delimited by full black (resp. dashed blue) lines. Metallic tubes, for which $(n-m)$ is a multiple of 3, are shown in brick red, and semi-conducting ones are flesh-colored. The parameter space for armchair (metallic) and $(n, n-1)$ and $(n, n-2)$ (semi-conducting) tubes is larger than for other chiralities.

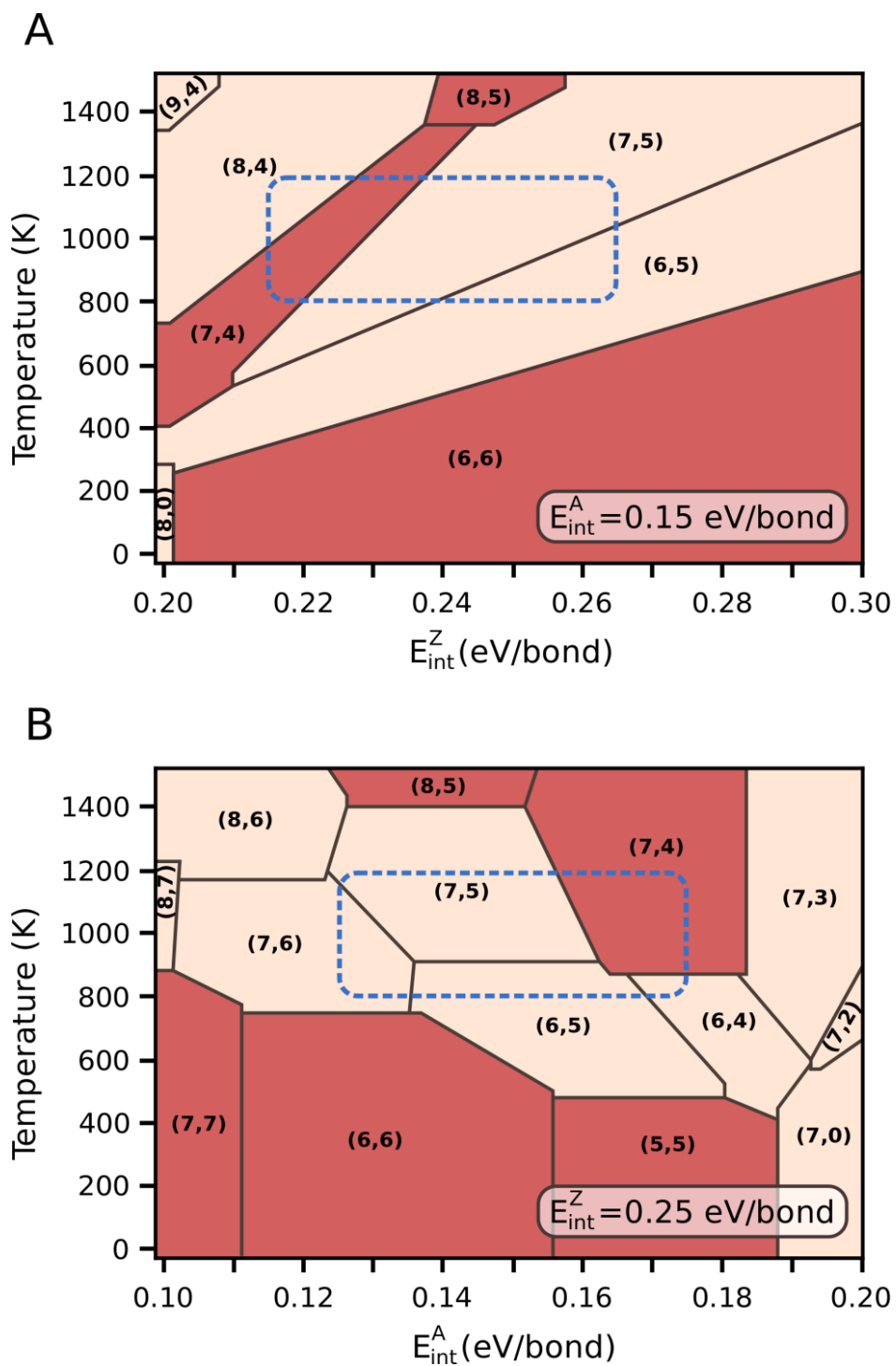

**Fig. 4. Chirality phase diagrams**. Phase diagrams calculated for constant values of $E_{Int}^{A}$ (**A**) and $E_{Int}^{Z}$ (**B**). These diagrams would be orthogonal in a 3-dimensional plot. The blue dashed boxes indicate possible parameter ranges corresponding to the analysis of growth products by He *et al. (26)*, based on a photoluminescence assignment of tubes grown using a FeCu catalyst. (6,5) tubes are reported stable up to 1023 K, (7,5) and (8,4) become dominant at 1023 K, and (7,6) at 1073 K.